\pdfoutput=1
\newcommand*{\ATLASLATEXPATH}{latex/}
\documentclass[cernpreprint, texlive=2016, UKenglish]{\ATLASLATEXPATH atlasdoc}
\usepackage{\ATLASLATEXPATH atlaspackage}
\usepackage{float}
\floatstyle{plaintop}
\restylefloat{table}

\usepackage{\ATLASLATEXPATH atlasbiblatex}
\usepackage{\ATLASLATEXPATH atlascontribute}

\usepackage{\ATLASLATEXPATH atlasphysics}

\usepackage[utf8]{inputenc}

\graphicspath{{logos/}{figures/}}

\usepackage{STDM-2017-15-PAPER-defs}

\AtlasTitle{Study of the hard double-parton scattering contribution to inclusive four-lepton production in $pp$ collisions at $\sqrt{s} = 8$ $\mathrm{TeV}$ with the ATLAS detector}

\AtlasAbstract{%
The inclusive production of four isolated charged leptons in $pp$ collisions is analysed for the presence of 
hard double-parton scattering, using $20.2$ $\mathrm{fb}^{-1}$ of data recorded in the ATLAS detector at the LHC at 
centre-of-mass energy $\sqrt{s} = 8$ $\mathrm{TeV}$.
In the four-lepton invariant-mass range of $80<m_{4\ell}<1000$ $\mathrm{GeV}$, an artificial neural network is used 
to enhance the separation between single- and double-parton scattering based on  the kinematics of the four leptons in 
the final state. An upper limit on the fraction of events originating from double-parton scattering is 
determined at 95\% confidence level to be $f_{\mathrm{DPS}} = 0.042$, which results in  an estimated lower limit 
on the effective cross section at 95\% confidence level of $1.0$ $\mathrm{mb}$.
}

\AtlasRefCode{STDM-2017-15}

\PreprintIdNumber{CERN-EP-2018-274}

\AtlasJournalRef{\PLB 790 (2019) 595}
\AtlasDOI{10.1016/j.physletb.2019.01.062}

\AtlasCoverSupportingNote{Double parton scattering with four-lepton final state at 8~TeV}{https://cds.cern.ch/record/2279957}

\AtlasCoverEgroupEditors{atlas-stdm-2017-15-contact-editors@cern.ch}

\AtlasCoverEgroupEdBoard{atlas-stdm-2017-15-editorial-board@cern.ch}

\hypersetup{pdftitle={ATLAS document},pdfauthor={The ATLAS Collaboration}}

\begin{document}

\maketitle
\tableofcontents

\section{Introduction}
\label{sec:intro}

The parton--parton scattering at the origin of hard processes in $pp$ interactions is accompanied by proton-remnant fragments that contribute to the hadronic final state through the so-called underlying event. 
As first pointed out by Sj{\"o}strand and van Zijl~\cite{Sjostrand:1987su}, one source of the 
underlying-event activity, particularly in the high-energy regime of the LHC, is multi-parton interactions (MPI): interactions of pairs of partons 
from the interacting protons which occur simultaneously with the hard process. 
In high-energy $pp$ interactions, where the density of low-$x$ partons is high, there is enough energy to produce 
hard multi-parton interactions.
The simplest example is hard double-parton scattering (DPS), 
where two partons from each proton interact with each other leading to perturbative final states.

The interest in studying DPS is twofold. Firstly, the probability of occurrence of DPS and the potential correlations between 
the products of these two perturbative interactions provide valuable information about the dynamics of the partonic structure of
the proton (see Ref.~\cite{Diehl:2017wew} and references therein). 
Secondly, DPS processes may also constitute a background to reactions proceeding through single-parton scattering (SPS). 
An example is the production of four charged leptons in the final state, addressed in this Letter. 
This reaction is dominated by the SPS production of two $Z^{(*)}$ bosons, followed by subsequent leptonic decays. 
The $Z^{(*)}$ notation indicates the production of on- or off-shell $Z$ bosons ($Z$ and $Z^{*}$), or the 
production of off-shell photons ($\gamma^{*}$). 
However,
the four leptons could also be produced as the result of two Drell--Yan processes occurring simultaneously, 
potentially distorting the measurements of prompt-lepton production. 

For a process $pp \rightarrow A + B + X$, the expected DPS cross section for producing states 
$A$ and $B$ in two independent scatterings, $\sigma_{\mathrm{DPS}}^{AB}$, may be estimated from the following 
formula~\cite{Diehl:2011tt,Diehl:2011yj,Manohar:2012jr} (see also Ref.~\cite{Bartalini:2011jp} for a detailed derivation):
\begin{equation}
\label{eq:pocket_formula}
\sigma_{\mathrm{DPS}}^{AB} = \frac{k}{2} \frac{\sigma_{\mathrm{SPS}}^{A} \sigma_{\mathrm{SPS}}^{B}} {\sigma_{\mathrm{eff}}},
\end{equation}
where $\sigma_{\mathrm{SPS}}^{A(B)}$ denotes the production cross section of state $A(B)$ in a single-parton scattering, 
the symmetry factor $k$ depends on whether the two scatterings lead to the same final state ($A=B$, $k=1$) or different 
final states ($A\neq B$, $k=2$), and $\sigma_{\mathrm{eff}}$ represents the effective transverse overlap area containing 
the interacting partons.

For most of the existing 
measurements~\cite{Akesson:1986iv,Alitti:1991rd,Abe:1993rv,Abe:1997xk,Abazov:2009gc,Aaij:2012dz,Aad:2013bjm,Chatrchyan:2013xxa,Aaboud:2016dea,Abazov:2014fha,Aad:2014kba,Aaij:2015wpa,Abazov:2015nnn,Aaij:2016bqq,Sirunyan:2017hlu}, $\sigma_{\mathrm{eff}}$ fluctuates around $15 \units{mb}$.
However, for the associated production of quarkonia $J/\psi J/\psi$ or $J/\psi \Upsilon$, $\seff$ is systematically lower~\cite{Abazov:2014qba,Abazov:2015fbl,BPHY-2015-02,Khachatryan:2016ydm} than for all 
other investigated processes. 
This might indicate that $\sigma_{\mathrm{eff}}$ is not universal and that there are spatial fluctuations of the 
parton densities in the proton, which may favour certain final states over others~\cite{Krasny:2013aca,Krasny:2015dka}. 
The concept of geometric fluctuations in the spatial parton densities has also been invoked~\cite{Mantysaari:2016ykx} to explain the 
collective phenomena observed in high-multiplicity proton--proton and 
proton--nucleus interactions~\cite{Khachatryan:2010gv,Abelev:2012ola,Aad:2012gla,Adare:2013piz}.
In $pp$ interactions at $\sqrt{s}=8~\TeV$~\cite{STDM-2014-15}, the double Drell--Yan contribution may add $0.3\%$ to the yield
of four leptons in the invariant-mass range $80~<~m_{4\ell}~<~1000~\GeV$, using Eq.~\eqref{eq:pocket_formula} 
with $\sigma_{\mathrm{eff}} = 15 \units{mb}$. 
The latter is obtained from calculations of the Drell--Yan cross section in the phase space of the measurement in 
next-to-leading-order (NLO) QCD with \POWHEGBOX~\cite{Melia:2011tj,Alioli2010,Nason:2004rx}.

Since double Drell--Yan production is driven by quark--antiquark annihilation, while most of the previously explored DPS processes are 
driven by gluon--gluon scattering, and the final state of four charged leptons constitutes the golden channel for the 
studies of Higgs boson properties, $H \rightarrow Z^{(*)}Z^{(*)} \rightarrow 4\ell$, 
a study of a possible DPS contribution to the production of four isolated charged leptons at $\sqrt{s}=8~\TeV$ is warranted. 
The analysis presented in this Letter closely follows a previous analysis of this final state~\cite{STDM-2014-15}, 
but extends it to consider DPS.

\section{ATLAS detector}
\label{sec:detector}
\newcommand{\AtlasCoordFootnote}{%
	ATLAS uses a right-handed coordinate system with its origin at the nominal interaction point (IP)
	in the centre of the detector and the $z$-axis along the beam pipe.
	The $x$-axis points from the IP to the centre of the LHC ring,
	and the $y$-axis points upwards.
	Cylindrical coordinates $(r,\phi)$ are used in the transverse plane, 
	$\phi$ being the azimuthal angle around the $z$-axis.
	The pseudorapidity is defined in terms of the polar angle $\theta$ as $\eta = -\ln \tan(\theta/2)$.
	Angular distance is measured in units of $\Delta R \equiv \sqrt{(\Delta\eta)^{2} + (\Delta\phi)^{2}}$.}

The ATLAS detector~\cite{PERF-2007-01} is a multipurpose particle detector with a forward--backward symmetric cylindrical geometry and 
nearly full coverage in solid angle.\footnote{\AtlasCoordFootnote}

 It consists of an inner tracking detector (ID) system surrounded by a superconducting solenoid, electromagnetic and hadronic calorimeters,
 and a muon spectrometer (MS) incorporating superconducting toroid magnets. During Run~1 of the LHC the ID consisted
 of a pixel detector closest to the beam-pipe, followed by a silicon strip detector and a transition radiation tracker. 
 This ID system, operating in a $2 \units{T}$ axial magnetic field, provides the tracking of charged particles within the 
 pseudorapidity range $|\eta| < 2.5$. 
 The calorimeter system, which covers the range $|\eta| < 4.9$, includes in the barrel region a high-granularity 
 lead/liquid-argon (LAr) barrel electromagnetic (EM) calorimeter ($|\eta| < 1.5$) and a steel/scintillator-tile hadronic calorimeter ($|\eta| < 1.7$). 
 In the endcap ($1.5 < |\eta| < 3.2 $) and forward ($3.2 < |\eta| < 4.9$) regions, the EM calorimeter and the hadronic calorimeter are made of LAr active layers with either copper or tungsten 
 as the absorber material.
 The muon spectrometer constitutes the outermost detector and includes fast trigger chambers covering the region $|\eta| < 2.4$ and 
 high-precision tracking chambers covering $|\eta| < 2.7$. 
 A three-level trigger system~\cite{PERF-2011-02} was used to select events to be recorded.

\section{Monte Carlo event samples}
\label{sec:mc_simulation}
In SPS events, the four-lepton events correspond to the production and subsequent decay of resonant $Z$ or Higgs bosons, or to the 
production of the continuum $Z^{(*)}Z^{(*)}$ system. 
In the case of DPS, the four leptons are decay products of two $Z^{(*)}$ bosons that are produced in two 
distinct parton--parton scatterings within the same $pp$ interaction.

The Monte Carlo samples are unchanged with respect to Ref.~\cite{STDM-2014-15}.
The SPS $q\bar{q} \rightarrow 4\ell$ was simulated with the \POWHEGBOX (revision 2330)~\cite{Melia:2011tj,Alioli2010,Nason:2004rx} Monte Carlo (MC)
program, which is based on perturbative QCD calculations
at NLO. The four-lepton production through the $qg$ initial state is included as part of the NLO contributions to the
$q\bar{q}$ process. The parton distribution functions (PDFs) of the CT10NLO~\cite{PhysRevD.82.074024} set were used.
The $gg \rightarrow 4\ell$ events corresponding to the continuum
$Z^{(*)}Z^{(*)}$ production were generated with \MCFM~6.1~\cite{Campbell:2013una} at leading order (LO) in QCD, using the CT10NNLO~\cite{PhysRevD.89.033009}
set of PDFs, and the cross sections were corrected for higher-order effects using the ratio of NLO to LO cross sections 
(the so-called $K$-factors)~\cite{Caola:2015psa}.
The on-shell Higgs boson production was simulated with \POWHEGBOX at NLO QCD, using the CT10NLO PDFs, in the case of 
gluon--gluon fusion and vector-boson fusion, and with LO \PYTHIAV 8~\cite{Sjostrand:2007gs} in the case of vector-boson 
associated production ($VH$) and top-pair associated production ($t\bar{t}H$). The event yield of on-shell Higgs boson was 
normalised to the higher-order corrected cross section~\cite{Heinemeyer:2013tqa}.
The events with off-shell Higgs boson production were simulated with the LO \MADGRAPH~5.1.5.12~\cite{Alwall:2014hca} generator via vector-boson fusion
and vector-boson scattering processes, including their interference. 
For the LO  \PYTHIAV 8 and \MADGRAPH generators, the LO version of CTEQ6L1 PDFs~\cite{1126-6708-2002-07-012} was used.

The MC generators listed above were interfaced to \PYTHIAV 8 for parton showering, except \MADGRAPH which was interfaced 
to \PYTHIAV 6~\cite{Sjostrand:2006za}. The underlying-event parameter values belong to the AU2~\cite{ATLAS:2012uec} tune. 

The DPS events that contribute to the $4\ell$ production were simulated with \PYTHIAV 8.175 using the LO version of CTEQ6L1 PDFs. 

Background events may originate from $Z + \mathrm{jets}$, $t\bar{t}$, diboson ($ZW$, $Z\gamma$), triboson $VVV$ ($V=Z,W$), $VH$, 
and $Z + \mathrm{top}$ ($t\bar{t}$ and $t$) processes.

The production of $Z+\mathrm{jets}$ events, including the light- and heavy-flavour contributions, was simulated 
with \ALPGEN~2.1.4~\cite{Mangano:2002ea}, using the Perugia2011C~\cite{Cooper:2011gk} tune. 
The $Z\gamma$ production was modelled with \SHERPA~1.4.5~\cite{1126-6708-2009-02-007}. 
Background $t\bar{t}$ events were generated with \POWHEGBOX using the Perugia2011C tune. 
The $ZH$ events, with subsequent decays $Z \rightarrow \ell \ell$  and $H \rightarrow VV^{*}$ (with two leptons
and two neutrinos or two leptons and two jets in the final state), were generated with \PYTHIAV 8, using the AU2 tune. 
The $ZW$ and $tZ$ processes were simulated with \SHERPA and \MADGRAPH respectively, with the latter 
using the AUET2B tune~\cite{ATLAS:2011zja}. The background contribution from $VVV$ and $t\bar{t}Z$ was modelled with \MADGRAPH, 
using the AUET2B tune. The MC generators for background simulation used the LO version of the CTEQ6L1 PDF set, except \SHERPA, 
which used the CT10 PDF set.

The largest contributions to the background, originating from $Z + b\bar{b}~\mathrm{jets}$ and $t\bar{t}$ production, were 
estimated in Ref.~\cite{STDM-2014-15} from the respective MC samples normalised to the data in selected control regions. 
The remaining background contributions were directly extracted from the MC expectations.

Additional $pp$ interactions occurring in the same and neighbouring bunch crossings (pile-up) 
were also simulated, using the \PYTHIAV 8 MC generator, with the A2~\cite{ATLAS:2014addedoleg} tune 
and MSTW 2008 LO~\cite{Sherstnev:2007nd} PDF set. The MC samples were reweighted to reproduce the distribution of the
mean number of $pp$ interactions per bunch crossing observed in the data.
The estimated number of events with two $Z^{(*)}$ bosons produced in the 
same bunch crossing with less than 1~cm separation along the beam axis is negligible compared to the DPS expectations.

Monte Carlo events were passed through the ATLAS detector simulation~\cite{SOFT-2010-01}, which is based on the \textsc{Geant4}~\cite{Agostinelli:2002hh} 
framework, and which includes simulation of the trigger selection. 
The MC events were reconstructed and selected offline using the same software and selections as for the data.

\section{Event selection}
\label{sec:event_selection}
The dataset and the event selection are unchanged with respect to Ref.~\cite{STDM-2014-15}. The updated luminosity of the analysed sample is $20.2$~$\mathrm{fb}^{-1}$.
The events were selected online using 
single-lepton or dilepton triggers. 
The single-lepton trigger required the transverse energy of the electron candidate or the transverse momentum of the muon 
candidate to be above $24~\GeV$. The dielectron trigger had the same threshold of $12~\GeV$ for both electron candidates. 
The dimuon trigger required either two muons with transverse momentum above $13~\GeV$ or one above $18~\GeV$ and the other above $8~\GeV$.
An electron--muon trigger was also used with thresholds at $12~\GeV$ for electrons and $8~\GeV$ for muons.

The final sample consists of events with at least four leptons, where each lepton is either an electron or a muon. 
The four leptons are required to form two same-flavour (electrons or muons) opposite-charge (SFOC) lepton pairs. 
The pair with the invariant mass closer to the mass of the $Z$ boson is called the leading pair, and the other pair is the sub-leading one. 
The invariant mass of the leading pair is restricted to the range $50 < m_{\mathrm{leading}} < 120~\GeV$, while for the sub-leading pair 
the mass requirement is $12 < m_{\textrm{sub-leading}} < 120~\GeV$. 
A $J/\psi$ veto is applied such that for any SFOC lepton combination the invariant mass of the dilepton, $m_{2\ell}$, must be 
greater than $5~\GeV$. Only events with the four-lepton invariant mass in the range $80 < m_{4\ell} < 1 000~\GeV$ are selected.  
The transverse momentum of dileptons, $\pT^{\ell^{+}\ell^{-}}$, is required to be above 2~GeV.  Selected leptons, ordered in descending 
order of transverse momentum, are required to have transverse momenta $\pT$ above 20, 15, 10 (8 if muon), and 7 (6 if muon)~GeV. 
The leptons are selected within the pseudorapidity
range $|\eta| < 2.5$ in the case of electrons and $|\eta| < 2.7$ in the case of muons.  
In order to have well-measured leptons, a lepton separation requirement is imposed, such that the distance between any two leptons 
in the $\eta$--$\phi$ space, $\Delta R$, is required to fulfil the condition $\Delta R > 0.1\,(0.2)$ for same-flavour (different-flavour) leptons.
Each event is required to have the triggering lepton(s) matched to one or two of the selected leptons.

\begin{figure}[htp]
      \begin{center}
              \centering
              \includegraphics[scale=0.5]{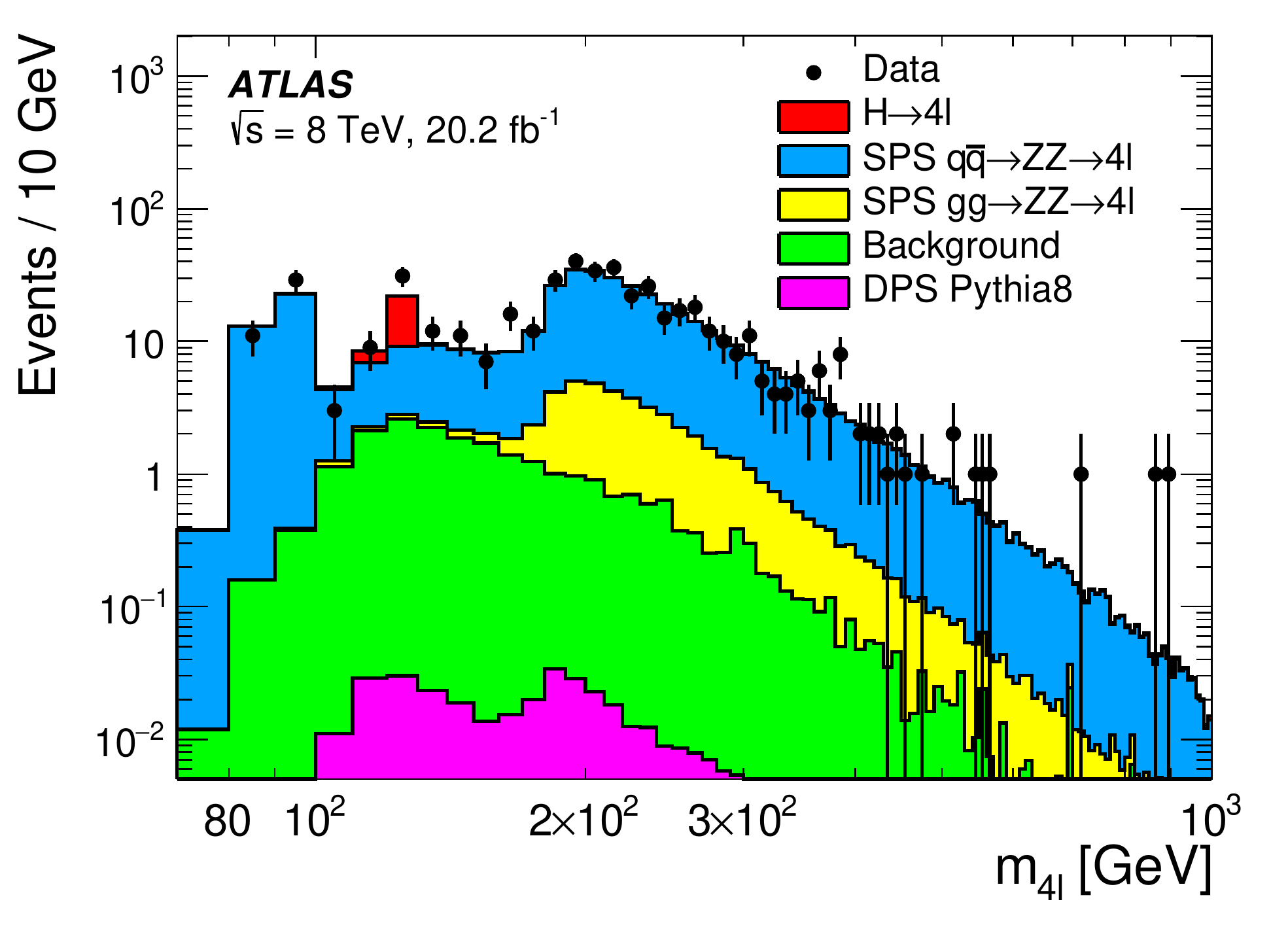}
              \caption{The distribution of the four-lepton invariant mass, $m_{4\ell}$. The data (black dots) are compared with the sum of signal and background MC expectations (filled coloured histograms). Also shown is the expected contribution of DPS from \PYTHIAV 8.}
              \label{fig:M4l}
      \end{center}
\end{figure}

The data sample, after all selections, contains 476 events. The resulting data and MC distributions of the four-lepton invariant mass are shown in 
Figure~\ref{fig:M4l}. 
For completeness, the figure also includes the DPS contribution of 0.4 events predicted by the \PYTHIAV 8.175 simulation.

\section{DPS signal extraction}
\label{sec:DPS_extraction}

The assumption that in DPS the two scatters are distinct implies that, in the DPS four-lepton final states, the two leptons of 
each dilepton will tend  to be balanced in \pT and therefore back-to-back in the azimuthal angle $\phi$, due to the dominance of 
low-\pT $Z^{(*)}$ production. In the SPS case, the leading and sub-leading pairs are expected to balance each other in $\pT$.

\begin{figure}[htp]
	\begin{center}
		\centering
		\hspace{0.0cm}\subfloat[]{\includegraphics[scale=0.35]{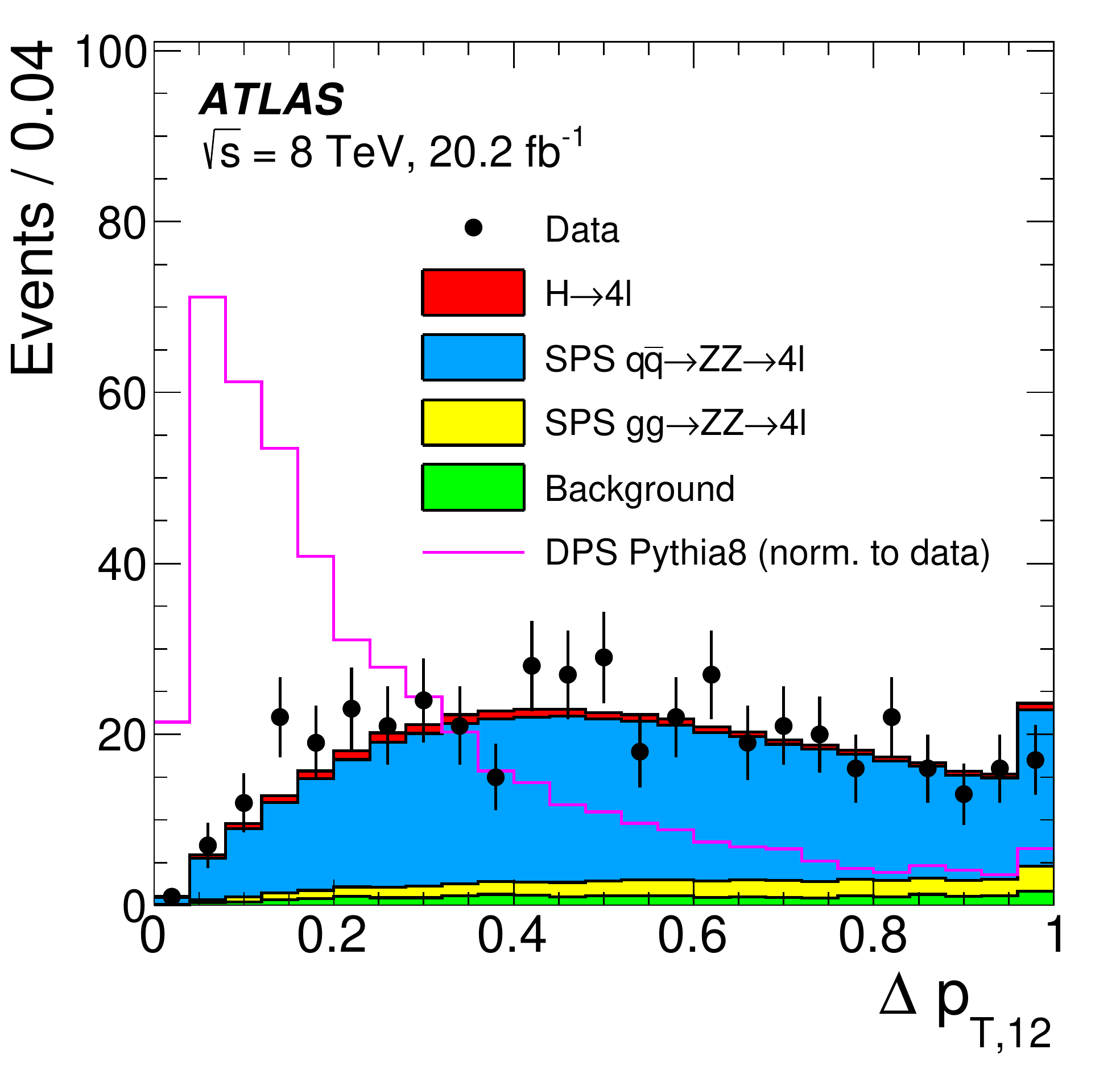}}
		\hspace{0.0cm}\subfloat[]{\includegraphics[scale=0.35]{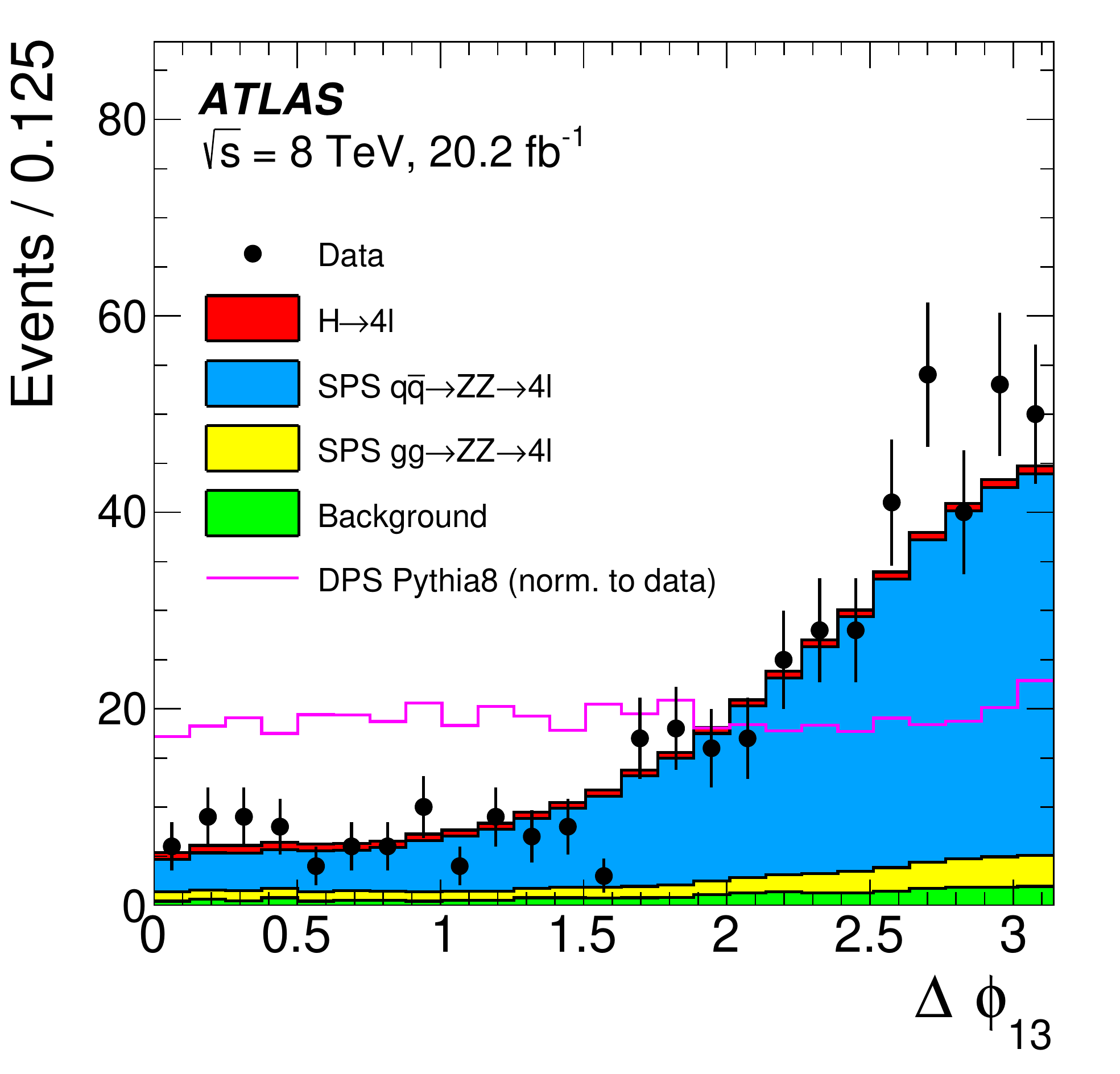}}\\
		\vspace{-0.3cm}
		\hspace{0.0cm}\subfloat[]{\includegraphics[scale=0.35]{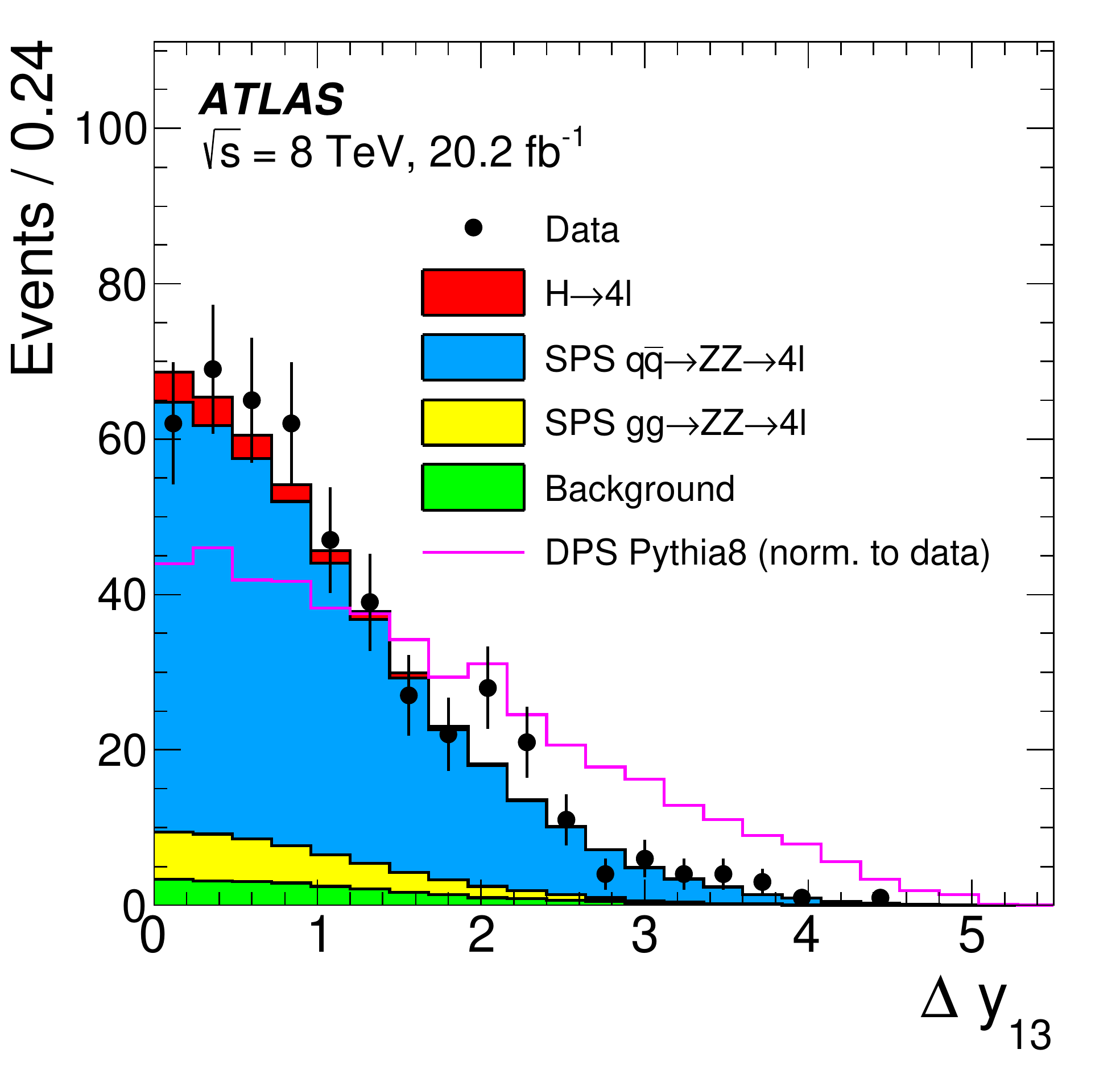}}
		\hspace{0.0cm}\subfloat[]{\includegraphics[scale=0.35]{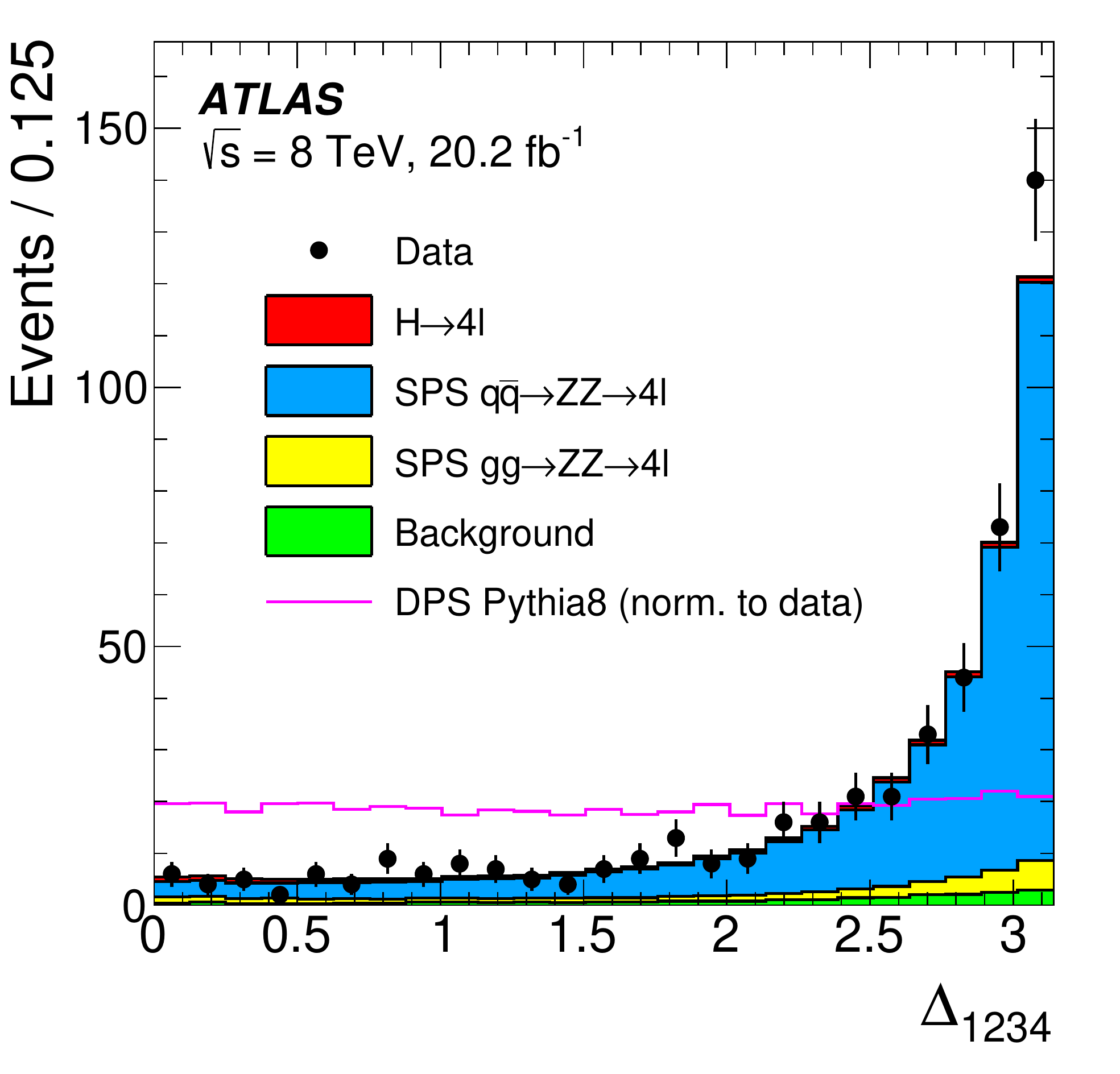}}
		
		\caption{Distributions of the discriminating variables (a) $\Delta p_{\mathrm{T},12}$, (b) $\Delta \phi_{13}$, (c) $\Delta y_{13}$, and (d) $\Delta_{1234}$. The definition of variables is given in Eq.~\eqref{eq:discriminating_variables}. Also plotted are the MC expectations for SPS and DPS, where the latter is normalised to the number of observed data events in order to make it clearly visible.}
		\label{fig:discr_variables}
	\end{center}
\end{figure}

Based on the experience gained in the study of four-jet final states~\cite{STDM-2015-08}, in order to distinguish between DPS events and SPS
events, the distributions of the following kinematic variables  of the four leptons are considered:

\begin{equation} \label{eq:discriminating_variables}
\begin{split}
& \Delta p_{\mathrm{T},ij} = \frac{\lvert \vec{p}_{\mathrm{T},i} + \vec{p}_{\mathrm{T},j} \rvert}{p_{\mathrm{T},i} + p_{\mathrm{T},j}}, \quad \Delta \phi_{ij} = \lvert \phi_i - \phi_j \rvert, \quad \Delta y_{ij} = \lvert y_i - y_j \rvert, \quad i,j=1,2,3,4,\quad i \neq j \\
& \Delta_{ijkm} = \lvert \phi_{i+j} - \phi_{k+m}  \rvert, \quad ijkm = 1234, 1324, 1423.
\end{split}
\end{equation}
Here, $\vec{p}_{\mathrm{T},i}$ is the transverse momentum component of the $i$-th lepton ($i = 1,2,3,4$), and $\phi_i$ and $y_i$ are the azimuthal 
angle and the rapidity of the $i$-th lepton, respectively. 
The angle $\phi_{i+j}$ is the azimuthal angle of the momentum vector composed by the sum of momenta of leptons $i$ and $j$.
Leptons 1 and 2 form the leading dilepton. The lepton ordering is chosen such that $p_{\mathrm{T,1}} > p_{\mathrm{T,2}}$ 
and $p_{\mathrm{T,3}} > p_{\mathrm{T,4}}$. 

The distributions of the variables $\Delta p_{\mathrm{T},12}$, $\Delta \phi_{13}$, $\Delta y_{13}$, and $\Delta_{1234}$ are presented 
in Figure~\ref{fig:discr_variables}(a)--(d). The distribution of $\Delta p_{\mathrm{T},12}$ peaks around 0.1 for simulated DPS events, 
while the simulated
SPS events are more evenly distributed across the range [0,1]. This demonstrates that, as expected, two leptons coming from the 
same $Z$ candidate in DPS balance each other in $\pT$, while in SPS the pairwise $\pT$ balance is not dominant. 
This is again demonstrated in  the $\Delta \phi_{13}$ distribution, where leptons 1 and 3 are decorrelated in $\Delta \phi$ for DPS, 
while for the SPS events these leading-\pT decay leptons tend to be back-to-back in $\phi$, because they originate from the 
two $Z$ bosons, which themselves are expected to be back-to-back in $\phi$. The $\Delta y_{13}$ distribution shows that leptons associated to different 
dileptons tend to be more separated in rapidity in DPS than in SPS. The back-to-back configurations of the two $Z$ candidates in 
the case of SPS, and their decorrelation in the case of DPS is explicitly demonstrated in the distribution of the azimuthal angle 
between two $Z$ candidates, $\Delta_{1234}$.

The difference between the topologies of SPS and DPS events is used to train an artificial neural network (ANN) to discriminate between the DPS and 
non-DPS classes, where the latter corresponds to SPS and background events.

The training is performed with the ANN available in the ROOT~\cite{Brun:1997pa} implementation of a feed-forward multilayer perceptron. 
The Broyden--Fletcher--Goldfarb--Shanno supervised learning 
algorithm~\cite{doi:10.1093/imamat/6.1.76,doi:10.1093/comjnl/13.3.317,Goldfarb:1970,10.2307/2004840} is used in the training. 
The input layer contains 21 neurons, corresponding to the variables listed in Eq.~\eqref{eq:discriminating_variables}, and the output 
layer consists of one neuron. As the result of optimising the convergence and the performance of the ANN, a configuration of 30 and 9 
neurons is adopted for the first and second hidden layer, respectively. 
The output of the ANN, $\xi_{\mathrm{DPS}}$, is a number distributed between 0 and 1, which represents the likelihood for an event to 
belong to the DPS class.

The event weights are chosen such that during the training procedure the effective numbers of 
SPS $q\bar{q}$-initiated events, $gg$-initiated events and background $Z + b\bar{b}~\mathrm{jets}$ events are in the ratio 
$1:1:1$. 
The SPS $gg$-initiated events tend to spill over into the DPS signal region, and a better separation between the SPS and DPS classes 
is achieved by increasing their weight in the minimisation of the error function. 
Similarly, the effective contribution of $Z + b\bar{b}~\mathrm{jets}$ events is increased for the ANN training  
to distinguish them better from the DPS ones, as the kinematics of the $Z + b\bar{b}~\mathrm{jets}$ background subprocess  
has features similar to DPS. The effective numbers of events for DPS and non-DPS events are equal.
Each MC set is split randomly into two subsets having approximately the same number of events. 
One subset is used for the ANN training, while the other is used to validate the performance of the ANN and to determine the 
number of training epochs, so as to reach the best possible level of discrimination while preventing overtraining.

\begin{figure}[htp]
	\begin{center}
		\centering
		\includegraphics[scale=0.5]{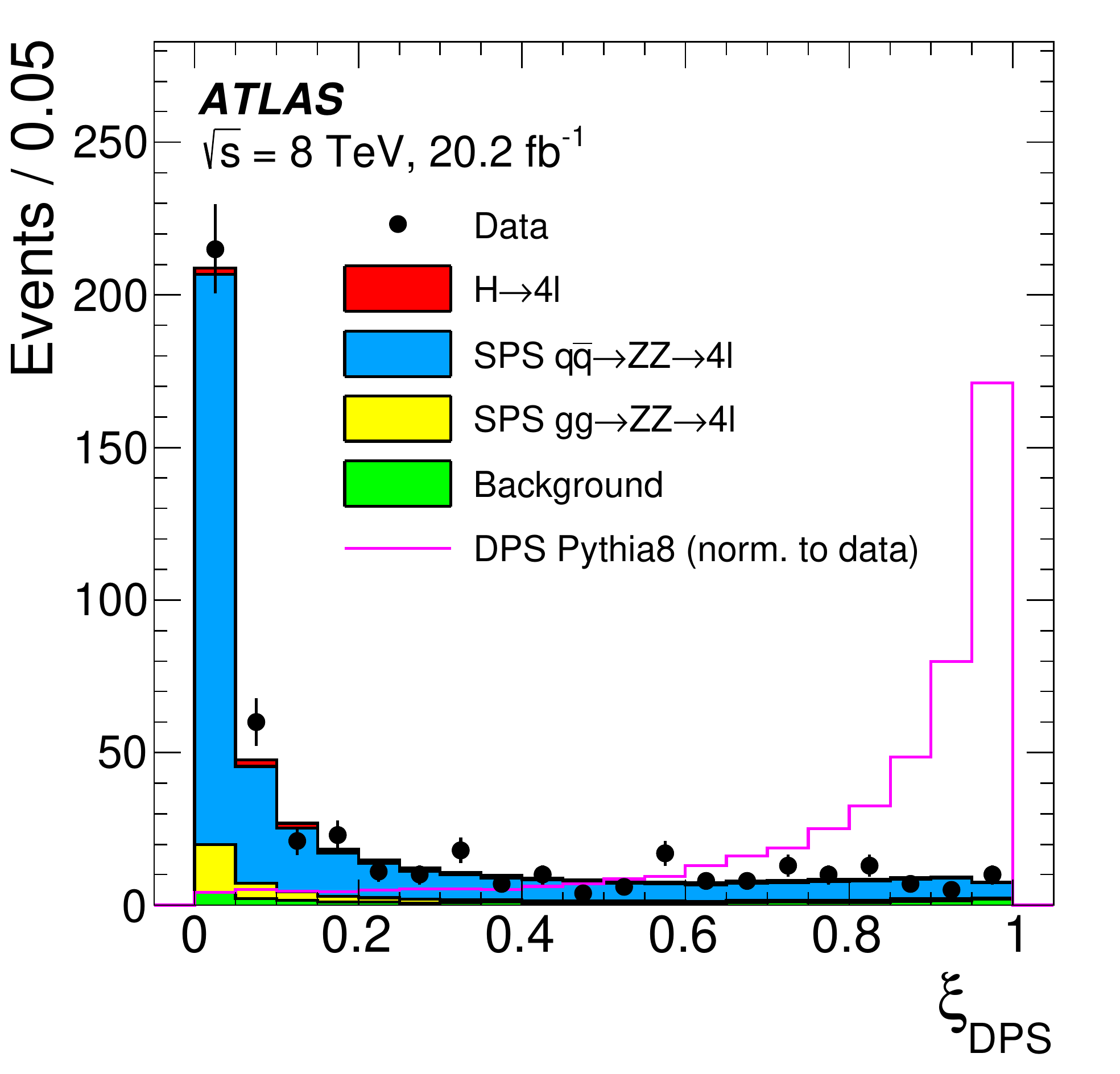}
		\caption{The distribution of the output variable of the artificial neural network, $\xi_{\mathrm{DPS}}$, shown 
		separately for the data, SPS, background, and DPS distributions.}
		\label{fig:NNdiscr}
	\end{center}
\end{figure}

The trained ANN is applied to data events, and the resulting distribution of  $\xi_{\mathrm{DPS}}$ is shown in Figure~\ref{fig:NNdiscr}, 
together with the corresponding DPS, SPS and background MC distributions. 
The DPS MC events form a peak around $\xi_{\mathrm{DPS}} = 1$ and the SPS and background events form a 
peak at $\xi_{\mathrm{DPS}} = 0$, as expected. A similar peak at $\xi_{\mathrm{DPS}} = 0$ is observed in data events, with no 
indication of a substantial contribution of double-parton scattering at $\xi_{\mathrm{DPS}} = 1$.

In order to quantify the level of the potential DPS contribution in the data, the variable $f_{\mathrm{DPS}}$ is introduced, 
defined as the ratio of the number of DPS events, $N_{\mathrm{DPS},4\ell}$, to the sum of the DPS and SPS ($N_{\mathrm{SPS},4\ell}$):
\begin{linenomath*}
\begin{equation*}
\label{eq:fDPS_definition}
f_{\mathrm{DPS}} = \frac{N_{\mathrm{DPS},4\ell}}{N_{\mathrm{SPS},4\ell}+N_{\mathrm{DPS},4\ell}}.
\end{equation*}
\end{linenomath*}
The MC template fit of the sum of the DPS, SPS and background contributions to the data 
yields $f_{\mathrm{DPS}} = \fDPSfitted \pm \fDPSfittedUnc$ with a $\chi^{2}$ per degree of freedom $\chi^2/\mathrm{dof} = 8.6 / 9$.
Since the result is consistent with zero, an upper limit on $f_{\mathrm{DPS}}$ is extracted, as 
described in Section~\ref{sec:upper_limit_fDPS}.

For the ANN performance to be robust and independent of the DPS model, it is best to have a DPS training sample with no inherent
correlations between the initial partons or the final states.
The DPS model in \PYTHIA~\cite{Sjostrand:2014zea,Corke:2010yf,Sjostrand:2017cdm} used in the analysis contains some correlations 
between the initial-state partons, implied by conservation of flavour and by the proton momentum sum-rule, as well as correlations 
due to inherent primordial transverse momentum of the partons and interleaved initial-state radiation. 
These effects are expected to be weak in the phase space of the present analysis (low-momentum partons and large transverse momenta 
of the final-state leptons). No correlations are expected in the production of the Drell--Yan final states.  

To test this assumption of a very weak correlation between two subscatterings in the \PYTHIA DPS model, the MC training sample was 
compared with a sample of two randomly overlaid dilepton events, where any correlation is eliminated by construction.
Such a sample was made by overlaying dilepton events selected 
in the data, with the selection driven by the four-lepton phase space. 
Each dilepton event was required to have two selected leptons forming an SFOC pair with transverse momenta $p_{\mathrm{T}}^{\ell_1, \ell_2} > 20,15~\GeV$ to 
account for the trigger conditions under which the dilepton data were collected. The same single-lepton, double-electron and 
double-muon triggers were used as in the selection of the four-lepton sample. An event was rejected if there was a third lepton 
with $p_{\mathrm{T}} > 7~\GeV$ ($6~\GeV$ for muons). 
The pairs of events were chosen randomly and overlaid by adding the lepton four-vectors of one event to the other. 
The distance between the primary vertices along the $z$-axis for the two events was required to be smaller than $1 \units{cm}$. 
After the overlay, the same four-lepton selection was applied as described in Section~\ref{sec:event_selection}, 
but the trigger configuration of the available dilepton datasets required an increase in the lepton $\pT$ thresholds. 
They were chosen to be 20, 20, 15, and 15~GeV for leptons ordered in descending order of \pT. 
To have a valid comparison within the same phase space between the overlaid dileptons and the \PYTHIAV 8 sample, the same selection
on lepton \pT 
was also applied to the latter. The distributions of discriminating variables were compared, as were the distributions 
of $\xi_{\mathrm{DPS}}$, obtained with the ANN trained on \PYTHIAV 8. 
Very good agreement between \PYTHIAV 8 and the overlaid data was observed, confirming the initial assumption of a very weak correlation
between the two scatterings in the \PYTHIA DPS model with no effect on the analysis.

The value of $f_{\mathrm{DPS}}$ is extracted using detector-level distributions. To test how well this result agrees with the 
parton-level value, $f_{\mathrm{DPS}}^{\mathrm{parton}}$, several pseudo-datasets were constructed by mixing DPS, SPS and background samples with a number of 
predefined parton-level values of $f_{\mathrm{DPS}}^{\mathrm{parton}}= 0.01, 0.03, 0.05, 0.1$, and 0.3. The number of background events
in all mixtures was the same as expected in the selected four-lepton data sample.
The corresponding value of $f_{\mathrm{DPS}}$ at the detector level was then determined by fitting the detector-level distributions
and compared with the input $f_{\mathrm{DPS}}^{\mathrm{parton}}$ value. It was found that the fitted value of $f_{\mathrm{DPS}}$ 
is systematically lower than $f_{\mathrm{DPS}}^{\mathrm{parton}}$ due to slightly different detector acceptances for DPS and SPS events. 
However, the two quantities agree within 2\%.

\section{Systematic uncertainties}
\label{sec:systematics}
The following sources of systematic uncertainty are considered:
\begin{itemize}
\item The experimental systematic uncertainty, which includes the uncertainties of the electron and muon energy scales, 
the uncertainty of the energy and momentum resolution, and of the trigger, reconstruction and identification 
efficiencies~\cite{PERF-2014-05,PERF-2013-05,PERF-2013-05}.
\item The uncertainty due to the model choice for the SPS process, which is evaluated by considering the effect of the variation 
of the fractions of $q\bar{q}$- and $gg$-initiated subprocesses, which are modelled with different MC generators, 
as described in Section~\ref{sec:mc_simulation}. 
For the determination of the range of variation, these fractions are fitted to the $m_{4\ell}$ distribution in the data, keeping the fraction of background events unchanged. 
The fraction values of $q\bar{q}$- and $gg$-initiated subprocesses were varied between the nominal values and the values obtained from the fit to the $m_{4\ell}$ distribution.
\item The uncertainty in the background modelling, which is estimated by varying the contributions of various background subprocesses  
according to the uncertainty of their normalisations obtained in Ref.~\cite{STDM-2014-15}
\end{itemize}

No uncertainty is assigned to the DPS model, since the kinematic distributions agree well between the \PYTHIAV 8 DPS model 
and the assumption of two independent interactions as represented by the overlaid dilepton data.

The combined effect of all systematic uncertainties, of which the variation of the $Z + b\bar{b}~\mathrm{jets}$ background is the dominant uncertainty, is about 20\%
of the statistical uncertainty on the fitted value of $f_{\mathrm{DPS}}$.
The effect of systematic uncertainties is therefore neglected when setting the upper limit on $f_{\mathrm{DPS}}$.

The validity of neglecting the systematic uncertainties was also checked with pseudo-experiments:
the contents of data bins were varied according to a Poisson distribution and those of MC profile histograms
were varied according to the systematic uncertainty, sampling the variations according to Gaussian distribution in the corresponding nuisance parameter,
taking into account the correlation between the bins where appropriate. 
For each set of varied data and MC histograms, the fit of  $f_{\mathrm{DPS}}$ was performed. 
The resulting distribution of $f_{\mathrm{DPS}}$ was compared with that obtained with systematic uncertainties neglected. 
The comparison showed no significant difference between the two distributions.

\section{Results}
\label{sec:result}
\subsection{Upper limit on $f_{\mathrm{DPS}}$}
\label{sec:upper_limit_fDPS}

The upper limit on $f_{\mathrm{DPS}}$ is determined using the distributions of the $\xi_{\mathrm{DPS}}$ variable in data, SPS, DPS, and
background MC samples. The statistical method to interpret the data uses the test statistic for upper limits, $q_{\mu}$, 
based on the profile likelihood ratio as described in Ref.~\cite{Cowan:2010js},
\begin{linenomath*}
\begin{equation*}
q_{\mu} = 
\begin{cases}
-2\ln \lambda(\mu) & \hat{\mu} \le \mu,\\
0 & \hat{\mu} > \mu.
\end{cases}
\end{equation*}
\end{linenomath*}
Here $\mu$ is the signal strength and $\lambda(\mu)$ is the profile likelihood ratio,
\begin{linenomath*}
\begin{equation*}
\lambda(\mu) = \frac{L(\mu, \hat{\hat{\theta}})}{L(\hat{\mu}, \hat{\theta})},
\end{equation*}
\end{linenomath*}
where $\theta$ is the number of non-DPS events and constitutes a nuisance parameter. 
The values $\hat{\mu}$ and $\hat{\theta}$ are maximum-likelihood estimators. 
The value of $\hat{\hat{\theta}}$ maximises $L$ for a given value of $\mu$.
 The parameter of interest, $\mu$, is defined to be equal to the $f_{\mathrm{DPS}}$ variable, $\mu = f_{\mathrm{DPS}}$. 
 Thus $\mu = 0$ corresponds to no DPS contribution, while $\mu = 1$ means that the 
 four-lepton sample consists exclusively of DPS events. 
 The procedure is that the data distribution is fitted with the sum of background, SPS and DPS histograms
 using the maximum-likelihood method. The upper limit is extracted using the $\mathrm{CL}_\mathrm{s}$ method~\cite{Read:2002hq} 
 from distributions of the test statistic for various hypothesised values of $\mu$. 
 The test-statistic distribution is obtained from an ensemble of pseudo-experiments.
 The shape of the test-statistic distribution agrees with the asymptotic formulae of Ref.~\cite{Cowan:2010js}. 
 The value of the $\mathrm{CL}_\mathrm{s}$ upper limit on $f_{\mathrm{DPS}}$ found with this method at 95\% confidence level (CL) 
 is \ULfDPS.

\subsection{Lower limit on the effective cross section}
\label{sec:lower_limit_sigma_eff}
The upper limit on $f_{\mathrm{DPS}}$ can be transformed into a lower limit on $\sigma_{\mathrm{eff}}$ by using 
Eq.~\eqref{eq:pocket_formula}. In order to perform this calculation, several inputs to the formula have to be determined.

The fiducial cross section for inclusive four-lepton production~\cite{STDM-2014-15} is
\begin{linenomath*}
\begin{equation*}
\sigma_{4\ell} = 32.0 \pm 1.6\, \mathrm{(stat.)} \pm 0.7\, \mathrm{(syst.)} \pm 0.9\, \mathrm{(lumi.)}~\mathrm{fb}.
\end{equation*}
\end{linenomath*}
The value of the symmetry factor $k/2$ in Eq.~\eqref{eq:pocket_formula} is well defined for the case of $2e + 2\mu$ or $2\mu + 2e$ final states, $k/2=1$. 
For the $4e$ or $4\mu$ final states, $k/2$ is well defined only in the case of completely overlapping ($k/2 = 1/2$) or fully 
exclusive ($k/2 = 1$) dilepton phase spaces. Therefore, the dilepton phase space is divided into 40 mutually exclusive regions.
The boundaries of these regions are driven by the lepton-\pT thresholds and by the dilepton 
invariant-mass ranges for the leading and sub-leading lepton pairs. 
The product $\frac{k}{2} \sigma_{A} \sigma_{B}$ is determined by representing Eq.~\eqref{eq:pocket_formula} as 
the sum over these phase-space regions. In order to determine the Drell--Yan cross section in each of the regions, 
the \POWHEGBOX MC simulation was used, based on NLO QCD calculations with the CT10 NLO set of PDFs.  
In the most populated region of $\pT >20 \units{GeV}$ for each lepton and of $50<m_{2\ell}< 120 \units{GeV}$, the calculated 
cross section is $0.55 \units{nb}$ for $2\mu$ and $0.49 \units{nb}$ for $2e$ final states. 
A conservative uncertainty of $\pm15\%$ is assigned to Drell--Yan cross sections. After summing the contributions from different 
dilepton phase-space regions, the result is
\begin{linenomath*}
\begin{equation*}
\frac{k}{2} \sigma_{A} \sigma_{B}  = (\prodDYsigma \pm \prodDYsigmaStat\, \mathrm{(stat)} \pm \prodDYsigmaSyst\, 
\mathrm{(syst)}) \cdot 10^{11}~\mathrm{fb}^2.
\end{equation*}
\end{linenomath*}
Here the systematic uncertainty is determined by propagating the assumed Drell--Yan cross-section uncertainty, 
assuming 100\% correlation between various phase-space regions.

\begin{figure}[htp]
	\begin{center}
		\centering
		\includegraphics[scale=1.0]{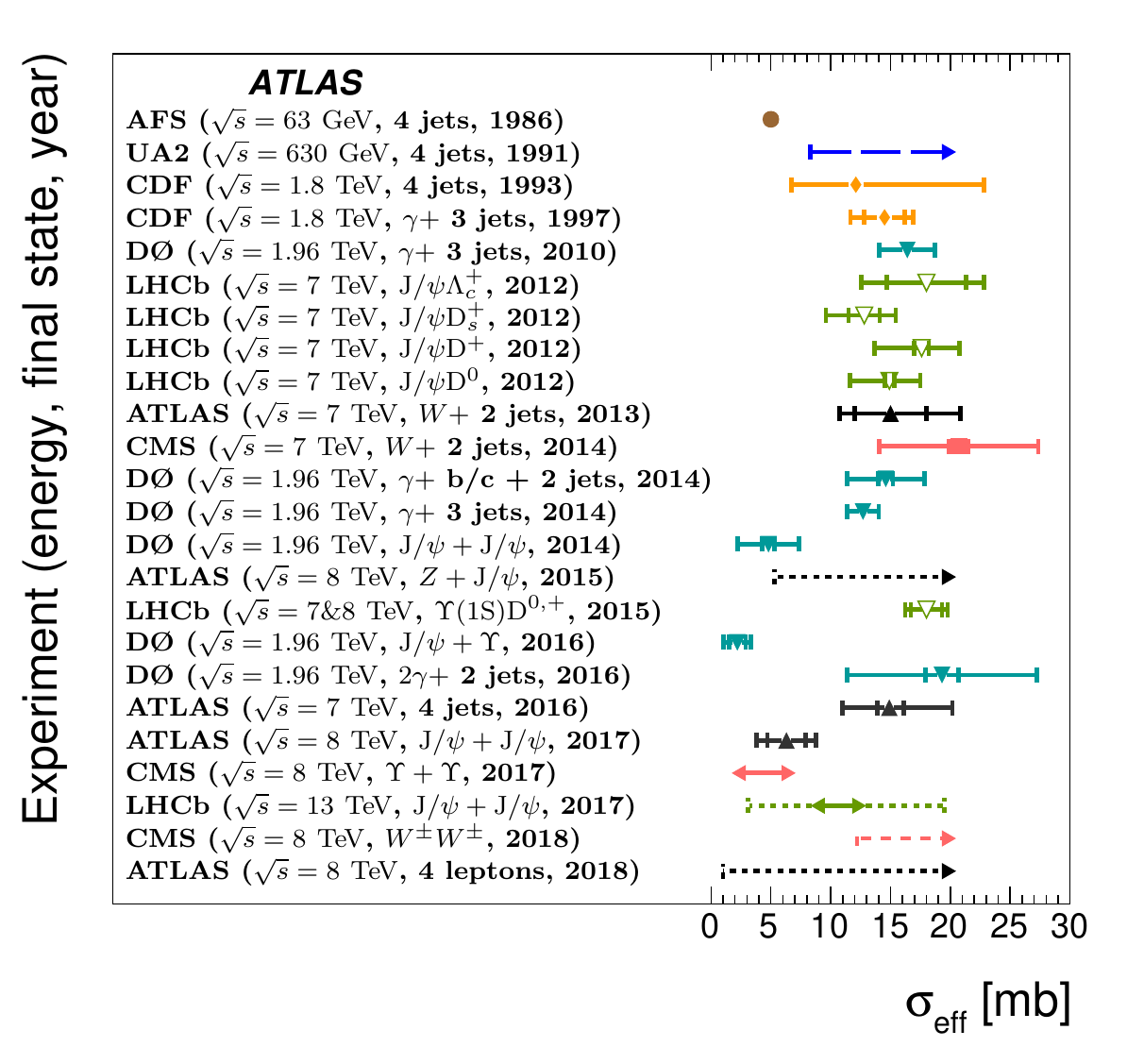}
		\caption{Summary of measurements and limits on the effective cross section, determined in different 
		experiments~\cite{Akesson:1986iv,Alitti:1991rd,Abe:1993rv,Abe:1997xk,Abazov:2009gc,Aaij:2012dz,Aad:2013bjm,
		Chatrchyan:2013xxa,Aaboud:2016dea,Abazov:2014fha,Abazov:2014qba,Aad:2014kba,Aaij:2015wpa,Abazov:2015nnn,Abazov:2015fbl,
		BPHY-2015-02,Khachatryan:2016ydm,Aaij:2016bqq,Sirunyan:2017hlu}, sorted chronologically. 
		The measurements that were made by different experiments are denoted by different symbols and colours. 
		The inner error bars represent statistical uncertainties and the outer error bars correspond to the total uncertainty. 
		Dashed arrows indicate lower limits. Lines with arrows on both ends represent ranges of the effective cross-section values, 
		determined within a single publication. In the case of the double $J/\psi$ measurement by LHCb, the dashed line denotes 
		the upper and lower uncertainties. The AFS measurement~\cite{Akesson:1986iv}, indicated with a dot, was 
		published without uncertainties.  }
		\label{fig:sigma_effective}
	\end{center}
\end{figure}

From the definition of $f_\mathrm{DPS}$, Eq.~\eqref{eq:pocket_formula} may be written as:
\begin{linenomath*}
\begin{equation*}
\frac{1}{\sigma_{\mathrm{eff}}} = \frac{f_{\mathrm{DPS}} \sigma^{4\ell} }{\frac{k}{2} \sigma^{A}_{\mathrm{SPS}}\sigma^{B}_{\mathrm{SPS}}},
\end{equation*}
\end{linenomath*}
and hence an approach similar to that used for the extraction of the upper limit on $f_\mathrm{DPS}$ can be applied to set the lower limit 
on $\sigma_{\mathrm{eff}}$.
The lower limit on $\sigma_{\mathrm{eff}}$ at 95\% CL is $\ULsigmaEff~\mathrm{mb}$, consistent with previously measured values of the 
effective cross section, as shown in Figure~\ref{fig:sigma_effective}.

\section{Summary}
\label{sec:conclusion}
The production of four-lepton (electrons or muons) final states in $pp$ interactions at $8~\TeV$ is analysed for the presence of 
double-parton scattering, using $20.2$~$\mathrm{fb}^{-1}$ of data recorded by the ATLAS experiment at the LHC.
Leptons with transverse momentum above 20, 15, 10 (8 if muon), and 7 (6 if muon)~$\GeV$, sorted 
in descending order of $\pT$, are selected in the pseudorapidity range $|\eta| < 2.5$ in the case of electrons and $|\eta| < 2.7$ in the case of muons. 
The four leptons form two same-flavour opposite-charge lepton pairs. The dilepton invariant masses are required to be 
in the range $50 < m_{\mathrm{leading}}  < 120~\GeV$ for the leading pair and $12 < m_{\textrm{sub-leading}} < 120~\GeV$ for the 
sub-leading pair, where the leading pair is defined as the pair with invariant mass closer to the $Z$ boson mass. 
The transverse momentum $\pT^{\ell^{+}\ell^{-}}$ of the dileptons is required to be above 2~GeV. 
The events in the four-lepton invariant-mass range of $80 < m_{4\ell} < 1 000~\GeV$ are considered. 
An artificial neural network is used to discriminate between single- and double-parton scattering events. 
No signal of double-parton scattering is observed and an upper limit on the fraction of the DPS contribution to the inclusive 
four-lepton final state of $\ULfDPS$ is obtained at 95\% CL. This upper limit translates, for two independent subscatterings, 
into a lower limit of $\ULsigmaEff \units{mb}$ on the effective cross section, consistent with previously measured values in 
different processes and at different centre-of-mass energies.

\section*{Acknowledgements}

We thank CERN for the very successful operation of the LHC, as well as the
support staff from our institutions without whom ATLAS could not be
operated efficiently.

We acknowledge the support of ANPCyT, Argentina; YerPhI, Armenia; ARC, Australia; BMWFW and FWF, Austria; ANAS, Azerbaijan; SSTC, Belarus; CNPq and FAPESP, Brazil; NSERC, NRC and CFI, Canada; CERN; CONICYT, Chile; CAS, MOST and NSFC, China; COLCIENCIAS, Colombia; MSMT CR, MPO CR and VSC CR, Czech Republic; DNRF and DNSRC, Denmark; IN2P3-CNRS, CEA-DRF/IRFU, France; SRNSFG, Georgia; BMBF, HGF, and MPG, Germany; GSRT, Greece; RGC, Hong Kong SAR, China; ISF and Benoziyo Center, Israel; INFN, Italy; MEXT and JSPS, Japan; CNRST, Morocco; NWO, Netherlands; RCN, Norway; MNiSW and NCN, Poland; FCT, Portugal; MNE/IFA, Romania; MES of Russia and NRC KI, Russian Federation; JINR; MESTD, Serbia; MSSR, Slovakia; ARRS and MIZ\v{S}, Slovenia; DST/NRF, South Africa; MINECO, Spain; SRC and Wallenberg Foundation, Sweden; SERI, SNSF and Cantons of Bern and Geneva, Switzerland; MOST, Taiwan; TAEK, Turkey; STFC, United Kingdom; DOE and NSF, United States of America. In addition, individual groups and members have received support from BCKDF, CANARIE, CRC and Compute Canada, Canada; COST, ERC, ERDF, Horizon 2020, and Marie Sk{\l}odowska-Curie Actions, European Union; Investissements d' Avenir Labex and Idex, ANR, France; DFG and AvH Foundation, Germany; Herakleitos, Thales and Aristeia programmes co-financed by EU-ESF and the Greek NSRF, Greece; BSF-NSF and GIF, Israel; CERCA Programme Generalitat de Catalunya, Spain; The Royal Society and Leverhulme Trust, United Kingdom. 

The crucial computing support from all WLCG partners is acknowledged gratefully, in particular from CERN, the ATLAS Tier-1 facilities at TRIUMF (Canada), NDGF (Denmark, Norway, Sweden), CC-IN2P3 (France), KIT/GridKA (Germany), INFN-CNAF (Italy), NL-T1 (Netherlands), PIC (Spain), ASGC (Taiwan), RAL (UK) and BNL (USA), the Tier-2 facilities worldwide and large non-WLCG resource providers. Major contributors of computing resources are listed in Ref.~\cite{ATL-GEN-PUB-2016-002}.

\printbibliography

\clearpage  
 
\begin{flushleft}
{\Large The ATLAS Collaboration}

\bigskip

M.~Aaboud$^\textrm{\scriptsize 34d}$,    
G.~Aad$^\textrm{\scriptsize 99}$,    
B.~Abbott$^\textrm{\scriptsize 125}$,    
O.~Abdinov$^\textrm{\scriptsize 13,*}$,    
B.~Abeloos$^\textrm{\scriptsize 129}$,    
D.K.~Abhayasinghe$^\textrm{\scriptsize 91}$,    
S.H.~Abidi$^\textrm{\scriptsize 164}$,    
O.S.~AbouZeid$^\textrm{\scriptsize 39}$,    
N.L.~Abraham$^\textrm{\scriptsize 153}$,    
H.~Abramowicz$^\textrm{\scriptsize 158}$
% ...
\bigskip
\\

% ...
$^{13}$Institute of Physics, Azerbaijan Academy of Sciences, Baku; Azerbaijan.\\
% ...
$^{34}$$^{(a)}$Facult\'e des Sciences Ain Chock, R\'eseau Universitaire de Physique des Hautes Energies - Universit\'e Hassan II, Casablanca;$^{(b)}$Centre National de l'Energie des Sciences Techniques Nucleaires (CNESTEN), Rabat;$^{(c)}$Facult\'e des Sciences Semlalia, Universit\'e Cadi Ayyad, LPHEA-Marrakech;$^{(d)}$Facult\'e des Sciences, Universit\'e Mohamed Premier and LPTPM, Oujda;$^{(e)}$Facult\'e des sciences, Universit\'e Mohammed V, Rabat; Morocco.\\
% ...
$^{39}$Niels Bohr Institute, University of Copenhagen, Copenhagen; Denmark.\\
% ...
$^{91}$Department of Physics, Royal Holloway University of London, Egham; United Kingdom.\\
% ...
$^{99}$CPPM, Aix-Marseille Universit\'e, CNRS/IN2P3, Marseille; France.\\
% ...
$^{125}$Homer L. Dodge Department of Physics and Astronomy, University of Oklahoma, Norman OK; United States of America.\\
% ...
$^{129}$LAL, Universit\'e Paris-Sud, CNRS/IN2P3, Universit\'e Paris-Saclay, Orsay; France.\\
% ...
$^{153}$Department of Physics and Astronomy, University of Sussex, Brighton; United Kingdom.\\
% ...
$^{158}$Raymond and Beverly Sackler School of Physics and Astronomy, Tel Aviv University, Tel Aviv; Israel.\\
% ...
$^{164}$Department of Physics, University of Toronto, Toronto ON; Canada.\\
% ...
$^{a}$ Also at Borough of Manhattan Community College, City University of New York, NY; United States of America.\\
$^{b}$ Also at California State University, East Bay; United States of America.\\
$^{c}$ Also at Centre for High Performance Computing, CSIR Campus, Rosebank, Cape Town; South Africa.\\
$^{d}$ Also at CERN, Geneva; Switzerland.\\
$^{e}$ Also at CPPM, Aix-Marseille Universit\'e, CNRS/IN2P3, Marseille; France.\\
% ...
$^{*}$ Deceased

\end{flushleft}


\end{document}